\documentclass[amsmath, pra, aps, twocolumn, showpacs, 10pt, floats]{revtex4-1}
\usepackage{graphicx}
\usepackage{longtable}

\begin{document}     

\title{State-dependent potentials in a nanofiber-based two-color trap for cold atoms}
 
\author{Fam Le Kien}

\author{P. Schneeweiss}

\author{A. Rauschenbeutel} 

\affiliation{Vienna Center for Quantum Science and Technology, Institute of Atomic and Subatomic Physics, Vienna University of Technology, Stadionallee 2, 1020 Vienna, Austria}

\date{\today}

\begin{abstract}
We analyze the ac Stark shift of a cesium atom interacting with far-off-resonance guided light fields in the nanofiber-based two-color optical dipole trap realized by Vetsch \textit{et al.} [Phys. Rev. Lett. \textbf{104}, 203603, (2010)]. Particular emphasis is given to the fictitious magnetic field produced by the vector polarizability of the atom in conjunction with the ellipticity of the polarization of the trapping fields. Taking into account the ac Stark shift, the atomic hyperfine interaction, and a magnetic interaction, we solve the stationary Schr\"odinger equation at a fixed point in space and find Zeeman-state-dependent trapping potentials. In analogy to the dynamics in magnetic traps, a local degeneracy of these state-dependent trapping potentials can cause spin flips and should thus be avoided.
We show that this is possible using an external magnetic field. Depending on the direction of this external magnetic field, the resulting trapping configuration can still exhibit state-dependent displacement of the potential minima. In this case, we find nonzero Franck-Condon factors between motional states of the potentials for different hyperfine-structure levels and propose the possibility of microwave cooling in a nanofiber-based two-color trap.
\end{abstract}

\pacs{42.50.Hz, 37.10.Gh, 32.10.Dk,  32.60.+i, 31.15.ap}
\maketitle

\section{Introduction} 
\label{sec:introduction}

Optically trapped neutral atoms are among the prime candidates for storing and processing quantum information~\cite{Kimble08,Polzik10}. This approach greatly profits from the use of microscopic dipole traps which enable the manipulation of individual neutral atoms~\cite{Schlosser01}. Several technical approaches have been followed in order to obtain optical dipole traps~\cite{Grimm00} with a microscopic trapping volume. Among these are strongly focused free-space optical traps, traps based on evanescent light fields, and plasmonically enhanced optical trapping potentials~\cite{Chang09,Gullans12}.

Tightly confined light fields are known to differ considerably from simple light fields which are described correctly in the paraxial approximation. For example, when tightly focusing an initially linearly polarized laser beam, the latter acquires a longitudinal polarization component that gives rise to a complex polarization pattern in the focal region. Consequently, the laser beam is not correctly described by a transversally polarized electromagnetic wave anymore. Recent experiments with atoms in the motional ground state of a non-paraxial dipole trap~\cite{Regal12,Vuletic12} as well as atoms trapped in the evanescent field of an optical nanofiber~\cite{Rauschenbeutel10,Goban12} contributed to the discussion of the ground state coherence properties of atoms in tightly-confining optical potentials~\cite{Lacroute12,Ding12}. In this article, we concentrate on the theoretical description of optical potentials in a nanofiber-based trap. However, our findings have implications for other technical realizations of tightly-confining optical potentials, too. 
 
\begin{figure}
\begin{center}
  \includegraphics{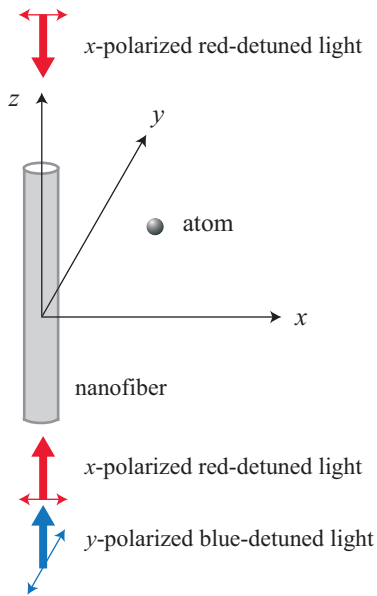}
 \end{center}
\caption{Nanofiber atom trapping scheme used in~\cite{Rauschenbeutel10}. The trapping potential is formed by combining the attraction and repulsion of  
red- and blue-detuned guided light beams, respectively. A pair of counter-propagating red-detuned beams (thick red arrows) and a single blue-detuned beam (thick blue arrow) are used. The red-detuned light beams are $x$-polarized (red thin arrows) and the blue-detuned light beam is $y$-polarized (thin blue arrow).}
\label{fig1}
\end{figure}

Among the diverse concepts of trapping using optical nanofiber-guided light fields~\cite{Dowling96,twocolors,Balykin04,Sague08}, only the so-called two-color trap has so far been realized experimentally~\cite{Rauschenbeutel10,Goban12} and will be discussed in this work. This trap confines atoms in the fiber transverse plane by combining the attraction and repulsion of a red- and blue-detuned guided light field~\cite{Dowling96,twocolors}. For axial confinement, the experiment of Vetsch \textit{et al.}~\cite{Rauschenbeutel10} used a pair of counter-propagating red-detuned beams and a single, running-wave blue-detuned beam (see Fig.~\ref{fig1}). The latter light field has an elliptical polarization which leads to a vector Zeeman-state-dependent light shift of the ground state of the trapped atoms~\cite{Lacroute12}. It has been proposed to replace the blue-detuned running wave by a rapidly moving blue-detuned standing wave field in order to create a ``compensated'' trap in which the ellipticity of the single blue-detuned field and thereby the vector light shift effect is reduced~\cite{Lacroute12}. This proposal has then been experimentally implemented in~\cite{Goban12}. Both versions of the two-color trap, the basic scheme and the compensated scheme have been compared theoretically in~\cite{Lacroute12}, in particular concerning the coherence times of ground state spin wave in trapped atomic ensembles. 

Here, we extend the discussion of state-dependent trapping potentials and the ground-state coherence of atoms in a two-color nanofiber-based optical potential of the type used in Vetsch \textit{et al.}~\cite{Rauschenbeutel10}. Most importantly, we point out deficiencies of the commonly used approach of \emph{adiabatic potentials} when calculating optical traps. As an example, we expect Majorana spin flips to play a role in optical traps that are at the same time steep and exhibit a strong gradient of the polarization of the trapping laser fields. Inspired from the physics of spin flips in magnetic traps for cold atoms, we study the influence of a magnetic offset field for atoms in the two-color nanofiber-based optical trap. Using the concept of light-induced fictitious magnetic fields, we show how an offset field allows one to modify the Zeeman-state-dependent potentials and to reduce spin flip effects. In addition, for a configuration that shows a Zeeman-state-dependent displacement of the trap minima, we point out the option to drive transitions between motional states of the atom in the trap using a microwave field. We calculate the Franck-Condon factors and suggest microwave cooling in a nanofiber-based optical potential.

The paper is organized as follows. In Sec.~\ref{sec:theory} we present the general formulae to calculate the ac Stark shift of a multilevel atom interacting with a far-off-resonance light field of arbitrary polarization and an external magnetic offset field. The concept to describe the influence of the nanofiber surface on the trapping potential is also discussed. In Sec.~\ref{sec:VetschTrap} we apply the formalism to describe the two-color nanofiber-based optical trap realized by Vetsch \textit{et al.}~\cite{Rauschenbeutel10} and present the results of numerical calculations for the adiabatic potentials of a cesium atom in the ground state $6S_{1/2}$ or the excited state $6P_{3/2}$. The deficiencies of the common approach of adiabatic potentials are discussed. In Sec.~\ref{sec:VetschMagnetic} we study the effect of an external magnetic field on the trap of Vetsch \textit{et al.}~\cite{Rauschenbeutel10} and find cases of non-zero Franck-Condon factors between motional states of different hyperfine-structure levels. Our conclusions are given in Sec.~\ref{sec:summary}.

\section{Formalism}
\label{sec:theory}

Let us examine the energy shifts of levels of a single fine-structure state. For this purpose, we will closely follow the formalism reviewed in Ref.~\cite{Fam12}. In general, due to the degeneracy of atomic levels and the possibility of level mixing, we must diagonalize the full interaction Hamiltonian in order to find the energy level shifts~\cite{Manakov86,Rosenbusch09}. In the case of an atom in a two-color nanofiber-based dipole trap, we have the following interaction Hamiltonian for the internal state $|nJ\rangle$ of the atom
\begin{equation}
H_{\mathrm{int}}=V^{\mathrm{hfs}}+V^{EE}+V^{B}.
\label{TotalHamiltonian}
\end{equation}
The atomic energy levels can be subject to perturbations by the Stark interaction $V^{EE}$, the hyperfine interaction (hfs) interaction $V^{\mathrm{hfs}}$, and the interaction with an external magnetic field $V^{B}$. For the description of the two-color trap, we take the hfs interaction from Ref.~\cite{Metcalf99} and discuss $V^{EE}$, $V^{B}$ as well as the interaction of the atom with the nanofiber surface in the following.

\subsection{ac Stark shift of a multilevel atom} 

Here, review the general theory of the ac Stark shift of a multilevel atom interacting with a far-off-resonance light field of arbitrary polarization~\cite{Manakov86,Mabuchi06,Jessen10,Rosenbusch09,Fam12}. 

In order to describe the internal states of the multilevel atom, we take a quantization axis $z_{Q}$, which can be, in general, arbitrary. We use this quantization axis
and an associated Cartesian coordinate frame $\{x_Q,y_Q,z_Q\}$ to specify bare basis states of the atom. Due to the hfs interaction, the total electronic angular momentum $\mathbf{J}$ is coupled to the nuclear spin $\mathbf{I}$. As a consequence, the projection $J_{z_Q}$ of the total electronic angular momentum $\mathbf{J}$ onto the quantization axis $z_Q$ is not conserved. However, in the absence of the external light field, the projection $F_{z_Q}$ of the total angular momentum $\mathbf{F}=\mathbf{J}+\mathbf{I}$ of the atom onto the quantization axis $z_Q$ is conserved.  We use the notation $|nJFM\rangle\equiv |nJFM\rangle_Q$ for the atomic hfs basis states, where $F$ is the quantum number for the total angular momentum $\mathbf{F}$ of the atom, 
$M$ is the quantum number for the projection $F_{z_Q}$ of $\mathbf{F}$ onto the quantization axis $z_Q$, 
$J$ is the quantum number for the total angular momentum $\mathbf{J}$ of the electron,
and $n$ is the set of the remaining quantum numbers $\{nLSI\}$, with $L$ and $S$ being the quantum numbers for the orbital angular momentum and the spin of the electrons, respectively.

Consider the interaction of the atom with a classical light field
\begin{equation}
\mathbf{E}=\frac{1}{2}\boldsymbol{\mathcal{E}}e^{-i\omega t}+\mathrm{c.c.}
=\frac{1}{2}\mathcal{E}\mathbf{u}e^{-i\omega t}+\mathrm{c.c.},
\label{n1}
\end{equation}
where $\omega$ is the angular frequency and $\boldsymbol{\mathcal{E}}=\mathcal{E}\mathbf{u}$ is the positive-frequency electric field envelope,
with $\mathcal{E}$ and $\mathbf{u}$ being the field amplitude and the polarization vector, respectively.
In general, $\mathcal{E}$ is a complex scalar and $\mathbf{u}$ is a complex unit vector. 

\subsubsection*{ac Stark interaction operator}

We assume that the light field is far from resonance with the atom. In addition, we take $J$ to be a good quantum number. This means that we treat only the cases where the Stark interaction energy is small compared to the fine structure splitting and, consequently, the level mixing of atomic states with different quantum numbers $J$ can be neglected. Then, the operator for the ac Stark interaction in the second-order perturbation theory with respect to the light field amplitude $\boldsymbol{\mathcal{E}}$ is given by~\cite{Manakov86,Rosenbusch09,Fam12}
\begin{eqnarray}
\lefteqn{V^{EE}= -\frac{1}{4}|\mathcal{E}|^2 \bigg\{\alpha^s_{nJ} 
-i\alpha^v_{nJ} \frac{[\mathbf{u}^*\times\mathbf{u}]\cdot\mathbf{J}}{2J}}
\nonumber\\&&\mbox{}
+\alpha^T_{nJ} \frac{3[(\mathbf{u}^*\cdot\mathbf{J})(\mathbf{u}\cdot\mathbf{J})
+(\mathbf{u}\cdot\mathbf{J})(\mathbf{u}^*\cdot\mathbf{J})]
-2\mathbf{J}^2}{2J(2J-1)}\bigg\}.
\label{n4}
\end{eqnarray}
The quantities $\alpha^s_{nJ}$,  $\alpha^v_{nJ}$, and $\alpha^T_{nJ}$ are the conventional dynamical scalar, vector, and tensor polarizabilities, respectively, of the atom in the fine-structure level $nJ$. They are given in~\cite{Manakov86,Rosenbusch09,Fam12}. Note that the tensor polarizability vanishes for $J=1/2$ states (e.g., the ground states of alkali-metal atoms). When the light is linearly polarized, the vector product $[\mathbf{u}^*\times\mathbf{u}]$ vanishes, making the contribution of the vector polarizability to the ac Stark shift to be zero. 

In the hfs basis $\{|nJFM\rangle\}$, the hfs interaction operator $V^{\mathrm{hfs}}$ is diagonal. 
However, the ac Stark interaction $V^{EE}$ is, in general, not diagonal in both indices $F$ and $M$ of the hfs basis. Therefore, in order to find the dressed eigenstates and eigenvalues of the atom, one has to diagonalize the total Hamiltonian (\ref{TotalHamiltonian}).

\subsubsection*{Fictitious magnetic fields} 
As the idea of fictitious magnetic fields will be of great importance to understand the nanofiber-based two-color trap, we will discuss this concept in detail. It is clear from Eq. (\ref{n4}) that the second term on the right-hand side of  this equation, which is responsible for the vector light shift,
can be written in a form that is similar to the operator for the interaction between a static magnetic field and an atom [see Eq. (\ref{n42})]. This means that the effect of the vector polarizability on the Stark shift is equivalent to that of a magnetic field with the induction vector~\cite{Cohen-Tannoudji72,Cho97,Zielonkowski98,Park01,Park02,Skalla95,Yang08,Rosatzin90,Wing84,Ketterle92,Kobayashi09,Miller02}  
\begin{equation}\label{n20} 
\mathbf{B}^{\mathrm{fict}}
=\frac{\alpha^v_{nJ}}{8\mu_Bg_{nJ}J} i[\boldsymbol{\mathcal{E}}^*\times\boldsymbol{\mathcal{E}}].
\end{equation}
Here, $\mu_B$ is the Bohr magneton and $g_{nJ}$ is the Land\'{e} factor for
the fine-structure level $nJ$.

The direction of the light-induced fictitious magnetic field $\mathbf{B}^{\mathrm{fict}}$ is determined by the vector 
$i[\boldsymbol{\mathcal{E}}^*\times\boldsymbol{\mathcal{E}}]$, which is a real vector.
Note that $\mathbf{B}^{\mathrm{fict}}$ is independent of $F$, that is,
$\mathbf{B}^{\mathrm{fict}}$ is the same for all hfs levels $nJF$ of a fine-structure level $nJ$.
If the light field is linearly polarized, we have $i[\boldsymbol{\mathcal{E}}^*\times\boldsymbol{\mathcal{E}}]=0$ and hence $\mathbf{B}^{\mathrm{fict}}=0$. 

In general, the vector Stark shift operator can be expressed in terms of the operator $\mathbf{J}$ as
$V^{EE}_{\mathrm{vec}}=\mu_Bg_{nJ}(\mathbf{J}\cdot\mathbf{B}^{\mathrm{fict}})$.
In the case of the ground state $nS_{1/2}$ of alkali-metal atoms,
since the hfs splitting of is very large compared to the light shift, the vector Stark shift operator can be given in terms of the operator $\mathbf{F}$ as
$V^{EE}_{\mathrm{vec}}=\mu_Bg_{nJF}(\mathbf{F}\cdot\mathbf{B}^{\mathrm{fict}})$.
Here, $g_{nJF}$ is the Land\'{e} factor for the hfs level $nJF$.
For the hfs levels $F=I\pm1/2$ of the ground state $nS_{1/2}$, we have $g_{nJF}|_{F=I\pm1/2}=\pm g_{nJ}/(2I+1)$. 
Hence, when the direction of the fictitious magnetic field $\mathbf{B}^{\mathrm{fict}}$ is taken as the quantization axis $z$, the vector Stark shifts of
the sublevels $M$ of the hfs levels $F=I\pm1/2$ of the ground state are
$V^{EE}_{\mathrm{vec}}|_{F=I\pm1/2;M}=\pm\mu_Bg_{nJ}MB^{\mathrm{fict}}/(2I+1)$.
These shifts are integer multiples of the quantity $\mu_Bg_{nJ}B^{\mathrm{fict}}/(2I+1)$. 
It is clear that the sublevels $M$ and $-M$ of the hfs levels $F=I+1/2$ and $F=I-1/2$, respectively, of the ground state have the same vector Stark shift.
In contrast, the sublevels with the same number $M$ of two different hfs levels $F=I\pm1/2$ have opposite vector Stark shifts. 

It is worth noting the similarities and differences between real and fictitious magnetic fields. 
Similar to a real magnetic field, the fictitious magnetic field is a pseudovector, that is,
$\mathbf{B}^{\mathrm{fict}}$ is invariant under space reflection. Both the real and fictitious magnetic fields change sign under time reversal. 
Both fields shift the magnetic sublevels of the atom in the same manner, expressed by the scalar product of the magnetic dipole of the atom and the 
magnetic induction vector of the field~\cite{Cohen-Tannoudji72,Cho97,Zielonkowski98,Park01,Park02,Skalla95,Yang08,Rosatzin90,Wing84,Ketterle92,Kobayashi09,Miller02}. 
Consequently, in calculations for level shifts, the fictitious magnetic field $\mathbf{B}^{\mathrm{fict}}$ can be vector-added to a real static magnetic field $\mathbf{B}$ if the latter is present in the system. Precession and echos of the atomic spin in an fictitious magnetic field have been demonstrated~\cite{Rosatzin90, Zielonkowski98}, and an optical Stern-Gerlach experiment has been performed~\cite{Park02}. Localized changes of the spin of an optically trapped ensemble as well as RF/MW-induced evaporative cooling of atoms in magnetic traps with contribution of fictitious magnetic fields are possible~\cite{Zielonkowski98,Miller02,Yang08}. 

There are limitations to the concept of fictitious magnetic fields~\cite{Cohen-Tannoudji72}. As the ratio $\alpha^v_{nJ}/g_{nJ}J$ is a quantity that depends on the considered atomic level $nJ$, so do the magnitude and sign of the fictitious magnetic field. This can be a restriction when the fictitious magnetic field is used as a quantization magnetic field for, e.g., optical pumping. Furthermore, for real magnetic fields, there can be no local maximum of the magnetic field in free-space~\cite{Wing84}. This restriction does not apply to fictitious magnetic fields: A circularly-polarized focussed Gaussian beam produces a fictitious magnetic field with a local maximum at the intensity maximum of the beam, which should allow one to build a ``magnetic'' trap for high-field-seeking paramagnetic atoms. In this context, the much faster switching time for optically induced fictitious magnetic fields as compared to real magnetic fields can turn out to be a technical advantage.

\subsection{Interaction with a static magnetic field}
The Hamiltonian for the interaction between the magnetic field and the atom is~\cite{Metcalf99}
\begin{equation}
V^B=\mu_Bg_{nJ} (\mathbf{J}\cdot\mathbf{B}).
\label{n42}
\end{equation}
It can be shown that the matrix elements of the operator $V^B$ in the basis $\{|FM\rangle\}$ are given by the expression
\begin{eqnarray}
V_{FMF'M'}^{B}&=&\mu_Bg_{nJ}(-1)^{J+I-M}
\sqrt{J(J+1)(2J+1)}
\nonumber\\&&\mbox{}
\times\sqrt{(2F+1)(2F'+1)}
\bigg\{\begin{array}{ccc}
F &1 &F' \\
J &I &J 
\end{array}\bigg\}
\nonumber\\&&\mbox{}
\times
\sum_{{q=0,\pm1}} 
(-1)^{q}B_{q}
\bigg(\begin{array}{ccc}
F &1 &F' \\
M & q& -M'
\end{array}\bigg).
\label{n44}
\end{eqnarray}
Here, $B_{-1}=(B_{x_Q}-iB_{y_Q})/\sqrt{2}$, $B_0= B_{z_Q}$, and $B_{1}= -(B_{x_Q}+iB_{y_Q})/\sqrt{2}$
are the spherical tensor components of the magnetic induction vector $\mathbf{B}=\{B_{x_Q},B_{y_Q},B_{z_Q}\}$.
We note that Eq. (\ref{n44}) is valid for an arbitrary quantization axis $z_Q$.

\begin{figure}
\begin{center}
  \includegraphics{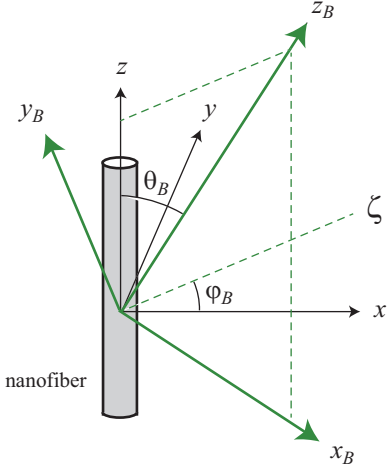}
 \end{center}
\caption{Orientation of the external magnetic field with respect to the nanofiber.
}
\label{fig2}
\end{figure} 

When $F$ is a good quantum number, the interaction operator (\ref{n42}) can be replaced by the operator
\begin{equation}
V^B=\mu_Bg_{nJF} (\mathbf{F}\cdot\mathbf{B}).
\label{n45}
\end{equation}
In the absence of the light field, the energies of the Zeeman sublevels are $\hbar\omega_{nJFM}=\hbar\omega_{nJF}+\mu_B g_{nJF}BM$. Here,  $\hbar\omega_{nJF}$ is the energy of the hfs level $nJF$ in the absence of the magnetic field and $M=-F,\dots,F$ is the magnetic quantum number. This  integer number is an eigenvalue corresponding to the eigenstate $|FM\rangle_B$ of the projection $F_{z_B}$ of $\mathbf{F}$ onto the $z_B$ axis. 

In general, the quantization axis $z_Q$ may be different from the magnetic field axis $z_B$ and, consequently, $|FM\rangle\equiv |FM\rangle_Q$ may be different from $|FM\rangle_B$. In order to find the level energy shifts, we must add the magnetic interaction operator $V^B$ to the combined hfs-plus-Stark interaction operator and then diagonalize the resulting operator~(\ref{TotalHamiltonian}). 

When we use the direction $z_B$ of the applied magnetic field as the quantization axis for the internal states of the atom (see Fig. \ref{fig2}), we need to describe the nanofiber-guided light field $\boldsymbol{\mathcal{E}}$ in a Cartesian coordinate system $\{x_B,y_B,z_B\}$. 
Let $\theta_B$ be the angle between the direction $z_B$ of the magnetic field and the fiber axis $z$.
Assume that the plane $(z,z_B)$ intersects with the fiber cross-section plane $(x,y)$ at a line $\zeta$. Let $\varphi_B$ be the angle between $\zeta$ and $x$. We choose the axes $x_B$ and $y_B$ such that $x_B$ is in the plane $(z_B,z)$ and $y_B$ is in the plane $(x,y)$. 
Then, the transformation for an arbitrary vector $\boldsymbol{\mathcal{E}}$ from the coordinate system $\{x,y,z\}$ to the coordinate system $\{x_B,y_B,z_B\}$ is given by the equations
\begin{eqnarray}
\mathcal{E}_{x_B}&=&(\mathcal{E}_x\cos\varphi_B+\mathcal{E}_y\sin\varphi_B)\cos\theta_B-\mathcal{E}_z\sin\theta_B,\nonumber\\ 
\mathcal{E}_{y_B}&=&-\mathcal{E}_x\sin\varphi_B+\mathcal{E}_y\cos\varphi_B,\nonumber\\  
\mathcal{E}_{z_B}&=&(\mathcal{E}_x\cos\varphi_B+\mathcal{E}_y\sin\varphi_B)\sin\theta_B+\mathcal{E}_z\cos\theta_B.\qquad
\label{n46}
\end{eqnarray}

\subsection{Atom-surface interaction}
\label{sec:surface}
We approximate the surface-induced potential by the van der Waals potential 
$U_{\mathrm{surf}}=-C_3/(r-a)^3$, where $C_3$ is the van der Waals coefficient for the interaction between an atom and a dielectric surface.
This approximation is justified by the fact that, in the close vicinity of the fiber surface, the geometry of the surface is not essential and, consequently, the atom sees the fiber surface as a flat surface~\cite{twocolors}. Meanwhile, in the region of large distances (comparable to or much larger than the fiber radius), the surface-induced potential is small and falls off faster than the optical potential~\cite{twocolors}. 
In the case considered here, the surface is silica and the atom is cesium. For numerical calculations, 
we take $C_3(6S_{1/2})=2\pi\hbar\times 1.16$ kHz $\mu$m$^3$~\cite{Kimble11} and 
$C_3(6P_{3/2})=2C_3(6S_{1/2})$~\cite{Ducloy,Courtois}.

\section{Nanofiber-based two-color trap}
\label{sec:VetschTrap}
In the experiment of Vetsch \textit{et al.}~\cite{Rauschenbeutel10}, atomic cesium was trapped in a nanofiber-based two-color optical potential~\cite{Dowling96,twocolors}. Two light fields, a red-detuned standing wave field and a blue-detuned running wave field, are launched into the fiber and are guided in the fundamental modes (see Fig. \ref{fig1}). The parameters of the experiment are the fiber radius $a=250$ nm, the wavelengths of the trapping fields $\lambda_1=1064$ nm (red-detuned with respect to the cesium D lines) and $\lambda_2=780$ nm (blue-detuned with respect to the cesium D lines), and the respective powers $P_1=2\times2.2$ mW and $P_2=25$ mW. 

\subsection{Dynamic polarizability of cesium}
\label{polarizability}
\begin{figure}
\begin{center}
  \includegraphics{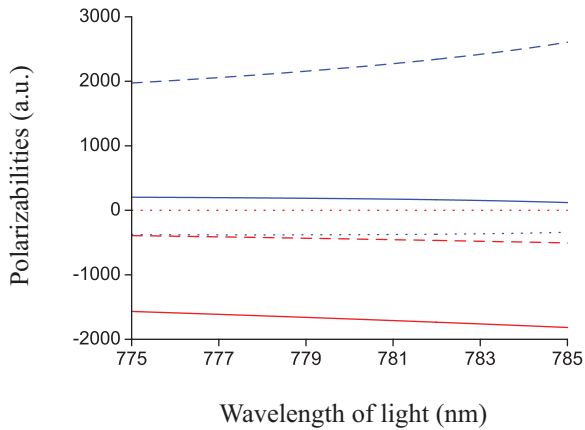}
 \end{center}
\caption{Polarizabilities of the ground state $6S_{1/2}$ (red color) and the excited state $6P_{3/2}$ (blue color) of atomic cesium in the region of wavelengths from 775 nm to 785 nm. The scalar, vector, and tensor components $\alpha^s_{nJ}$, $\alpha^v_{nJ}$, and  $\alpha^T_{nJ}$ are shown by the solid, dashed, and dotted curves, respectively.}
\label{fig3}
\end{figure}

\begin{figure}
\begin{center}
  \includegraphics{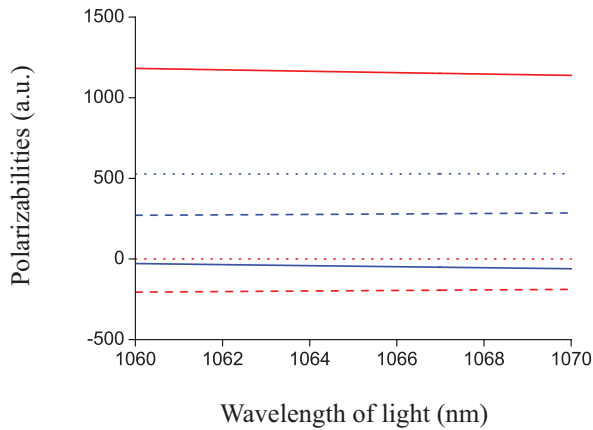}
 \end{center}
\caption{Same as Fig. \ref{fig3} but in the region of wavelengths from 1060 nm to 1070 nm.}
\label{fig4}
\end{figure}

For calculating the optical potentials of the two-color trap with the formalism presented in \ref{sec:theory}, the polarizability of cesium for the trapping wavelength is required. We plot in Figs. \ref{fig3} and \ref{fig4} the polarizabilities $\alpha^s_{nJ}$ (solid lines), $\alpha^v_{nJ}$ (dashed lines), and $\alpha^T_{nJ}$ (dotted lines) of the ground state $6S_{1/2}$ (red color) and the excited state  $6P_{3/2}$ (blue color) in the regions around the wavelengths $780$ nm and $1064$ nm of the trapping light fields. The calculation is based on the formalism and the data set provided in Ref.~\cite{Fam12}. We observe from Figs.~\ref{fig3} and \ref{fig4} that the magnitude of the vector and the tensor polarizabilities $\alpha^v_{nJ}$ $\alpha^T_{nJ}$ can be, in general, substantial. This means that the ac Stark shift can be different for each Zeeman state, depending on the polarization of the light field.

\subsection{Guided modes of optical nanofibers}
\label{subsec:guided}
When the radius of the nanofiber is well below the wavelengths of the guided light fields, the nanofiber supports only the hybrid fundamental modes HE$_{11}$ corresponding to each given wavelength~\cite{fiber books}. The light field in such a mode is strongly guided. It penetrates into the outside of the nanofiber in the form of an evanescent wave carrying a significant fraction of the optical power~\cite{fibermode}.
In order to describe guided light fields, we use Cartesian coordinates  $\{x,y,z\}$ and associated cylindrical coordinates $\{r,\varphi,z\}$, with $z$ being the fiber axis. 

Suppose, the nanofiber is a silica cylinder of radius $a$ and refractive index $n_1$ and is surrounded by an infinite vacuum of refractive index $n_2=1$. For a fundamental guided mode HE$_{11}$ of a light field of frequency $\omega$ (free-space wavelength $\lambda=2\pi c/\omega$ and free-space wave number $k=\omega/c$), the propagation constant $\beta$ is determined by the
fiber eigenvalue equation~\cite{fiber books}.
The cylindrical components of the unnormalized mode-profile vector function $\mathbf{e}(\mathbf{r})$ of the electric part of the fundamental guided mode are given, for $r>a$, by~\cite{fiber books}
\begin{eqnarray}
e_{r}&=&i[(1-s)K_0(qr)+(1+s)K_2(qr) ],
\nonumber\\
e_{\varphi}&=&-[(1-s)K_0(qr)-(1+s)K_2(qr) ],
\nonumber\\
e_{z}&=& \frac{2q}{\beta}K_1(qr).
\label{n25}
\end{eqnarray} 
Here the parameter $s$ is defined as
$s=({1}/{h^2a^2}+{1}/{q^2a^2})/[{J_1^\prime (ha)}/{haJ_1(ha)}+{K_1^\prime (qa)}/{qaK_1(qa)}]$. The parameters $h=(n_1^2k^2-\beta^2)^{1/2}$ and $q=(\beta^2-n_2^2k^2)^{1/2}$ 
characterize the fields inside and outside the fiber, respectively. The notations $J_n$ and $K_n$ stand for the Bessel functions of the first kind and the modified Bessel functions of the second kind, respectively. 
We note that the axial component $e_{z}$ is significant in the case of nanofibers~\cite{fibermode}. This makes guided modes of nanofibers very different from plane-wave modes of free-space and from guided modes of conventional, weakly guiding optical fibers~\cite{fibermode,fiber books}.

In the experiment by Vetsch \textit{et al.}~\cite{Rauschenbeutel10}, an optical potential with a trapping minimum sufficiently far from the fiber surface was produced by two guided light fields $\boldsymbol{\mathcal{E}}_1$ and $\boldsymbol{\mathcal{E}}_2$ in the fundamental modes with red- and blue-detuned frequencies $\omega_1$ and $\omega_2$, respectively. In the experiment, the field $\boldsymbol{\mathcal{E}}_1$ is a pair of counter-propagating $x$-polarized beams, while the field $\boldsymbol{\mathcal{E}}_2$ is a single $y$-polarized beam (see Fig.~\ref{fig1}). The combined field of the pair of counter-propagating $x$-polarized red-detuned beams is
\begin{eqnarray}
\boldsymbol{\mathcal{E}}_1&=&
A_1\{ [\hat{\mathbf{x}}(e_r\cos^2\varphi-ie_\varphi\sin^2\varphi)
+\hat{\mathbf{y}}(e_r+ie_\varphi)
\nonumber\\&&\mbox{}
\times\sin\varphi\cos\varphi]\cos\beta z
+i\hat{\mathbf{z}}e_z\cos\varphi\sin\beta z\}. 
\label{n36}
\end{eqnarray}
In deriving the above equation we have assumed, without loss of generality, that the point $z=0$ corresponds to an antinode of the transverse component of the field. The polarization of $\boldsymbol{\mathcal{E}}_1$ is linear at every point in space.

Meanwhile, the single $y$-polarized blue-detuned field is
\begin{eqnarray}
\boldsymbol{\mathcal{E}}_{2}
&=&A_2[\hat{\mathbf{x}}(e_r+i e_\varphi)\sin\varphi\cos\varphi
\nonumber\\&&\mbox{}
+\hat{\mathbf{y}} (e_r\sin^2\varphi
-i e_\varphi\cos^2\varphi)
+\hat{\mathbf{z}}e_z\sin\varphi] e^{i\beta z}.\qquad
\label{n37}
\end{eqnarray}

The above equations for the trapping light fields in conjunction with the atomic polarizability given in \ref{polarizability} are used to calculate the optical potential of the two-color trap.

\subsection{Adiabatic trapping potentials for a cesium atom}
\label{sec:potentials}
In this section, we present the results of numerical calculations for the adiabatic potentials of the ground state $6S_{1/2}$ and the excited state $6P_{3/2}$ of a cesium atom in the two-color nanofiber-based trap realized in the experiment~\cite{Rauschenbeutel10,Vetsch10}.

The total atomic trap potential $U$ consists of the optical potential $U_{\mathrm{opt}}$ and the surface-induced  potential $U_{\mathrm{surf}}$ (see \ref{sec:surface}), that is,
\begin{equation}\label{n69}
U=U_{\mathrm{opt}}+U_{\mathrm{surf}}.
\end{equation}
In the work by Vetsch \textit{et al.} \cite{Rauschenbeutel10}, there was no external magnetic offset field present and we therefore neglect such an interaction for the moment. The optical potential $U_{\mathrm{opt}}$ is produced by the light shifts of the atomic energy levels and obtained by diagonalizing the total interaction Hamiltonian (see \ref{sec:theory})
\begin{equation}\label{n70}
H_{\mathrm{int}}=V^{\mathrm{hfs}}+V^{EE}_R+V^{EE}_B
\end{equation}
at each point in space. Here, $V^{EE}_R$ and $V^{EE}_B$ are the operators for the Stark interaction caused by the red- and blue-detuned light fields, respectively. In view of the large mutual detuning, 
the interference between the red- and blue-detuned light fields has been neglected in Eq. (\ref{n70}).

\begin{figure}
\begin{center}
 \includegraphics{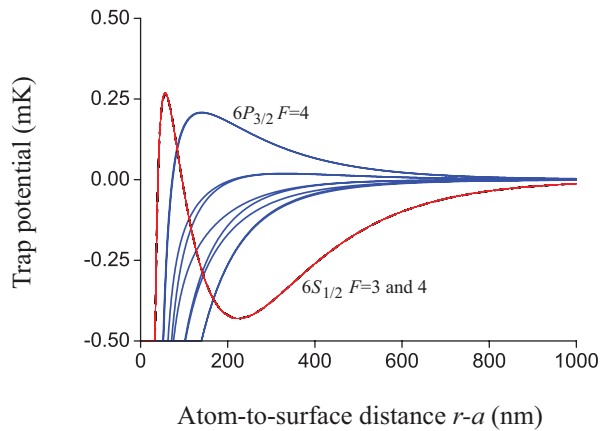}
 \end{center}
\caption{Radial dependence of the potentials of the ground and excited states of a cesium atom in the nanofiber-based trap.
The sublevels of the excited-state hfs level $6P_{3/2}\,F=4$ are shown as solid blue curves, and the sublevels of the ground-state hfs levels $6S_{1/2}\,F=3$ and $6S_{1/2}\,F=4$ are shown as dashed black and solid red curves, respectively. The azimuthal and axial coordinates of the atom are $\varphi=0$ and $z=0$, respectively. For parameters of the trap, see the text.}
\label{fig5}
\end{figure}

\begin{figure}
\begin{center}
 \includegraphics{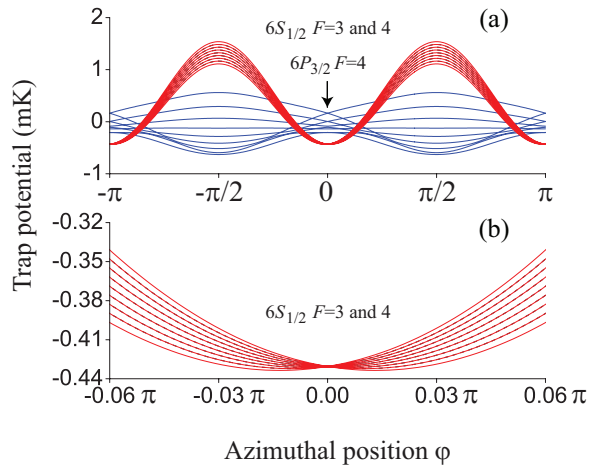}
 \end{center}
\caption{(a) Azimuthal dependence of the potentials of the ground and excited states of a cesium atom in the nanofiber-based trap. 
The sublevels of the excited-state hfs level $6P_{3/2}\,F=4$ are shown as solid blue curves, 
and the sublevels of the ground-state hfs levels $6S_{1/2}\,F=3$ and $6S_{1/2}\,F=4$ are shown as dashed black and solid red curves, respectively.
The radial distance and axial position of the atom are $r-a=224$ nm and $z=0$, respectively. For parameters of the trap, see the text.
(b) Expanded view of the potentials of the ground-state hfs levels in Fig. \ref{fig6}(a) around the trap minimum at $\varphi=0$. The sublevels of the ground-state hfs levels $6S_{1/2}\,F=3$ and $6S_{1/2}\,F=4$ are shown as dashed black and solid red curves, respectively.
}
\label{fig6}
\end{figure}

\begin{figure}
\begin{center}
 \includegraphics{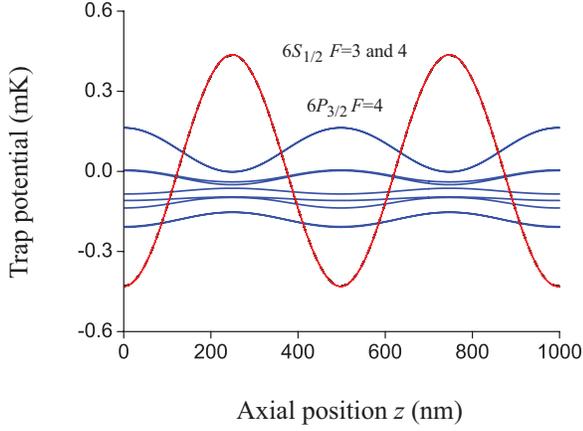}
 \end{center}
\caption{Axial dependence of the potentials of the ground and excited states of a cesium atom in the nanofiber-based trap. The sublevels of the excited-state hfs level $6P_{3/2}\,F=4$ are shown as solid blue curves, 
and the sublevels of the ground-state hfs levels $6S_{1/2}\,F=3$ and $6S_{1/2}\,F=4$ are shown as dashed black and solid red curves, respectively.
The radial distance and azimuthal position of the atom are $r-a=224$ nm and $\varphi=0$, respectively. For parameters of the trap, see the text. 
}
\label{fig7}
\end{figure}

We plot in Figs. \ref{fig5}, \ref{fig6}, and \ref{fig7} the radial, azimuthal, and axial dependences of the potentials 
of the sublevels of the hfs levels $F=3$ and $F=4$ of the ground state $6S_{1/2}$ and of the hfs level $F=4$ of the excited state $6P_{3/2}$ of a cesium atom in the nanofiber-based trap. 
In order to avoid overcrowding, we do not show the potentials of the other $6P_{3/2}$ hfs levels.
The dashed black and solid red curves of the figures, corresponding to the potentials for the sublevels of the hfs levels $F=3$ and $F=4$ of the ground state $6S_{1/2}$, respectively, show clearly that there are trap minima at the positions with the coordinates $r-a=224$ nm, $\varphi=0,\pm\pi$, and $z=0$. Since the red-detuned field is a standing wave along the fiber axis, there are two arrays of trapping minima~\cite{Rauschenbeutel10}. 
The trap depth is about $0.43$ mK, $1.75$ mK, and $0.87$ mK in the radial, azimuthal, and axial directions, respectively.
The blue curves, corresponding to the potentials for the sublevels of the hfs level $F=4$ of the excited state $6P_{3/2}$, show that the excited states are not trapped. 

The intersection of the dashed black and solid red curves in Fig. \ref{fig6}(a) occurs at $\varphi=0,\pm\pi$, that is, on the $x$ axis, where $B^{\mathrm{fict}}=0$. The details of the potentials of the ground-state hfs levels around the trap minimum at $\varphi=0$ are expanded in Fig. \ref{fig6}(b). 
We observe from Fig. \ref{fig6}  that the degeneracy of the sublevels of the hfs levels of the ground state is lifted at $\varphi\not=0,\pm\pi$ but remains at $\varphi=0,\pm\pi$, and the energy separation between these sublevels varies in the azimuthal direction. Such a behavior is a consequence of the azimuthal dependence of the magnitude of the vector Stark shift produced by the interaction with the quasi-linearly polarized running-wave blue-detuned light field.

We also see from Fig. \ref{fig6}(b) that the local minima of the potentials of the ground-state hfs sublevels in the azimuthal direction are slightly displaced from each other in the vicinities of the intersection points $\varphi=0$. The same displacements also occur in the vicinity of $\varphi=\pm\pi$. The reason is that, at $\varphi=0,\pm\pi$, the light-induced fictitious magnetic field $\mathbf{B}^{\mathrm{fict}}$ changes its direction from $+\hat{\mathbf{x}}$ to $-\hat{\mathbf{x}}$ (see Fig.\ \ref{fig8}), and so do the differential shifts $\mu_Bg_{nJF}B^{\mathrm{fict}}(M-M')$ of the sublevels $|FM\rangle_{B^{\mathrm{fict}}}$
and  $|FM'\rangle_{B^{\mathrm{fict}}}$. Here, $|FM\rangle_{B^{\mathrm{fict}}}$ are the eigenstates of the angular momentum projection operator $F_{z_{B^{\mathrm{fict}}}}$ with the direction of  $\mathbf{B}^{\mathrm{fict}}$ taken as the quantization axis. 
We observe from Figs. \ref{fig5},  \ref{fig6}, and  \ref{fig7} that the sublevels $M$ and $-M$ of the hfs levels $F=4$ and $F=3$, respectively, of the ground state have the same vector Stark shift and, consequently, the same potential. Meanwhile, 
the sublevels with the same quantum number $M\not=0$ of the different hfs levels $F=4$ and $F=3$ have opposite vector Stark shifts
and, consequently, oppositely displaced potentials.

\subsection{Limits of the concept of adiabatic potentials}
\label{sec:potential}
As outlined, we diagonalize the interaction Hamiltonian (\ref{n70}) at fixed points ${\mathbf{R}}$ in space. Then, we get a set of local eigenstates $|F({\mathbf{R}})M({\mathbf{R}})\rangle$ 
and local eigenvalues $E_{FM}(\mathbf{R})$. We subtract the hfs splittings from the obtained eigenvalues $E_{FM}(\mathbf{R})$ and add the van der Waals potential. As a result, we get a set of potential branches $U_{FM}(\mathbf{R})$, which are called adiabatic potentials.

Such adiabatic potentials are not conventional potentials for the translational motion~\cite{Berry84,Aharonov92,Littlejohn93,Sukumar97,Gov00,Brink06}. 
The reason is that the Hamiltonian (\ref{n70}) acts on both the internal state and the center-of-mass motion of the atom.
Because of this, the diagonalization of the Hamiltonian (\ref{n70}) with respect to only the internal degrees of freedom of the atom is just a partial diagonalization. The full Hamiltonian for the atom is $H=T+H_{\mathrm{int}}$, where $T=\mathbf{P}^2/2m$ is the operator for the kinetic energy, with $\mathbf{P}$ and $m$ being the momentum and the mass of the atom, respectively. In terms of the local internal eigenstates $|F({\mathbf{R}})M({\mathbf{R}})\rangle$, we can write $H_{\mathrm{int}}=\sum_{FM} E_{FM}(\mathbf{R}) |F(\mathbf{R})M(\mathbf{R})\rangle \langle F(\mathbf{R})M(\mathbf{R})|$. We introduce the unitary transformation $H'=\mathcal{R}^{-1}(\mathbf{R})H\mathcal{R}(\mathbf{R})$ and $|\Psi'\rangle=\mathcal{R}^{-1}(\mathbf{R})|\Psi\rangle$ for the full Hamiltonian $H$ and the full state vector $|\Psi\rangle$ of the atom, where $\mathcal{R}(\mathbf{R})$ is a position-dependent unitary operator. Such a transformation is similar to the use of a position-dependent coordinate system for describing a spin in an inhomogeneous magnetic field~\cite{Aharonov92,Littlejohn93}. We choose $\mathcal{R}(\mathbf{R})=\sum_{FMF'M'} \langle F'M'|F({\mathbf{R}})M({\mathbf{R}})\rangle |F'M'\rangle\langle FM|$, which acts on the internal states such that $|F({\mathbf{R}})M({\mathbf{R}})\rangle=\mathcal{R}(\mathbf{R})|FM\rangle$. When we perform the above transformation, we obtain $H'=H_{\mathrm{ad}}+\Delta T$, where $H_{\mathrm{ad}}=T+\sum_{FM} E_{FM}(\mathbf{R}) |FM\rangle \langle FM|$ is the adiabatic Hamiltonian and $\Delta T=\mathcal{R}^{-1}(\mathbf{R})T\mathcal{R}(\mathbf{R})-T=[\mathbf{A}(\mathbf{R})\cdot\mathbf{P}+\mathbf{P}\cdot\mathbf{A}(\mathbf{R})+\mathbf{A}^2(\mathbf{R})]/2m$ is the nonadiabatic correction, with $\mathbf{A}(\mathbf{R})=-i\hbar \mathcal{R}^{-1}(\mathbf{R}) \nabla  \mathcal{R}(\mathbf{R})$~\cite{Sukumar97,Brink06}. The omission of $\Delta T$ constitutes the adiabatic approximation. 
In this approximation, the local internal eigenstate of the atom is rigidly coupled to the translational degrees of freedom and, consequently, the atom may be considered as having only translational degrees of freedom. 
Under this approximation, the atom follows a potential branch $U_{FM}(\mathbf{R})$ to remain in an adiabatic state 
$|\Psi(t)\rangle=|F(\mathbf{R})M(\mathbf{R})\rangle\psi(\mathbf{R},t)$ if it is initially prepared in a local internal eigenstate $|F(\mathbf{R})M(\mathbf{R})\rangle$. Here, $\psi(\mathbf{R},t)$ is the time-dependent wave function for the center-of-mass motion of the atom in the potential $U_{FM}(\mathbf{R})$. The adiabatic approximation holds true when the matrix elements 
$\langle F'M'|F(\mathbf{R})M(\mathbf{R})\rangle$ vary in space slowly enough that $\Delta T$ can be considered as a small perturbation. 

Like in the case of magnetic traps, two effects may arise when the nonadiabatic term $\Delta T$ is not negligible:  
(1) the atom can undergo Majorana  spin-flip transitions~\cite{Majorana32} (due to nondiagonal matrix elements of $\Delta T$) and (2) the motion of the atom in an adiabatic potential is modified by geometric forces~\cite{Berry84,Aharonov92,Littlejohn93}
(due to diagonal matrix elements of $\Delta T$). For magnetic traps, the rates of Majorana transitions have been calculated  by a perturbation method for $F= 1/2$~\cite{Sukumar97,Gov00} and $F=1$~\cite{Gov00} and for $F>1$~\cite{Brink06}, and the geometric forces have been discussed~\cite{Berry84,Aharonov92,Littlejohn93}.

To illustrate the situations where the concept of adiabatic potentials becomes problematic, let us reconsider the potential in the two-color trap. In Fig.~\ref{fig6}, the results of the diagonalization of the Hamiltonian at each point in space is shown in dependence of the azimuthal coordinate. As apparent from the plot, there is a degeneracy at $\varphi=0,\pm\pi$. This indicates that the adiabatic approximation is not valid in this situation. Because of this breakdown, the internal and external dynamics of an atom with a finite center-of-mass motion cannot be described by the calculated adiabatic potentials. However, due to the dominant effect of the scalar Stark shift, all adiabatic potential branches of the ground state are trapping potentials for the atom. Therefore, if its kinetic energy is low enough, the atom remains in the trap even when undergoing spin-flip transitions.

The study of spin-flip transitions and geometric forces is beyond the scope of this paper. However, in order to reach long ground-state coherence times of the nanofiber-trapped atom, e.~g.,~for electromagnetically induced transparency (EIT) experiments or the implementation of a quantum memory, spin-flips of the trapped atoms have to be suppressed.

These effects have not been considered in the analysis of the experiment of Vetsch \textit{et al.}~\cite{Rauschenbeutel10} by the authors of Ref.~\cite{Lacroute12}. However, we have experimental evidence that they have to be taken into account when estimating ground-state coherence times \cite{Reitz13}.

\subsubsection{Fictitious magnetic fields by nanofiber-guided light fields}
\label{subsec:guided_fictitious}

In section \ref{sec:theory}, we introduced the scalar, vector, and tensor polarizabilities for the description of the interaction of the atom with a light field and pointed out that the tensor polarizability is zero $J=1/2$ states. Consequently, for the ground state of a cesium atom in the two-color trap will be only subject to the scalar and the vector light shift. The scalar light shift is the same for all Zeeman states and, consequently, leads to Zeeman-state-independent trapping potentials. The situation is different for the contribution of the vector light shift to the trapping potential which can be interpreted as a fictitious magnetic field (see Sec.~\ref{sec:theory}). In order to obtain a deeper understanding of the potentials presented in Sec.~\ref{sec:potentials}, we calculate the fictitious magnetic fields that appear in the two-color trap.

Let us consider the following two specific cases: a quasi-linearly polarized running-wave mode and a pair of counter-propagating modes with the same quasi-linear polarization. This is the light field configuration used by Vetsch \textit{et al.}~\cite{Rauschenbeutel10} and illustrated in Fig.~\ref{fig1}. The field in a quasi-linearly polarized running-wave mode can be composed as a superposition of the
left- and right-handed circular fields, that is,
\begin{eqnarray}
\boldsymbol{\mathcal{E}}_{\mathrm{lin}}
&=&A[\hat{\mathbf{r}}e_r\cos(\varphi-\varphi_0)
+i\hat{\boldsymbol{\varphi}}e_\varphi\sin(\varphi-\varphi_0)
\nonumber\\&&\mbox{}
+f\hat{\mathbf{z}}e_z\cos(\varphi-\varphi_0)] 
e^{if\beta z}.
\label{n28}
\end{eqnarray}
Here the angle $\varphi_0$ specifies the principal direction of the polarization vector in the fiber transverse plane, and
the index $f=1$ or $-1$ stands for the forward or backward propagation direction, respectively.

In the case of quasi-linear polarization, the field is given by
\begin{eqnarray}
\boldsymbol{\mathcal{E}}_{\mathrm{lin}\mbox{-}\mathrm{st}}
&=&A[\hat{\mathbf{r}}e_r\cos(\varphi-\varphi_0)\cos\beta (z-z_0)
\nonumber\\&&\mbox{}
+i\hat{\boldsymbol{\varphi}}e_\varphi\sin(\varphi-\varphi_0)\cos\beta (z-z_0)
\nonumber\\&&\mbox{}
+i\hat{\mathbf{z}}e_z\cos(\varphi-\varphi_0)\sin\beta (z-z_0)] .
\label{n30}
\end{eqnarray}
The parameter $z_0$ specifies the positions of the nodes and antinodes along the fiber axis.

We now study the fictitious magnetic field $\mathbf{B}^{\mathrm{fict}}$ produced by the light field in a nanofiber guided mode. As discussed in the previous section, 
the vector Stark shift of an atom is equivalent to that caused by a fictitious magnetic field $\mathbf{B}^{\mathrm{fict}}$ oriented along the direction of the vector $i[\boldsymbol{\mathcal{E}}^*\times\boldsymbol{\mathcal{E}}]$~\cite{Cohen-Tannoudji72,Cho97,Zielonkowski98,Park01,Park02,Skalla95,Yang08,Rosatzin90,Wing84,Ketterle92,Kobayashi09,Miller02}. 
In general, we have
\begin{equation}\label{n31}
i[\boldsymbol{\mathcal{E}}^*\times\boldsymbol{\mathcal{E}}]=2\mathrm{Im}(\mathcal{E}_\varphi\mathcal{E}_z^*)\hat{\mathbf{r}}
+2\mathrm{Im}(\mathcal{E}_z\mathcal{E}_r^*)\hat{\boldsymbol{\varphi}}+2\mathrm{Im}(\mathcal{E}_r\mathcal{E}_\varphi^*)\hat{\mathbf{z}}.
\end{equation}
Note that the spatial dependences of $\mathcal{E}_r$, $\mathcal{E}_\varphi$, and $\mathcal{E}_z$ are determined by the mode profile functions $e_r$, $e_\varphi$, and $e_z$ and the polarization of the field.
Unlike plane-wave light fields, a guided light field may have a nonzero axial component $\mathcal{E}_z$.
Therefore, the fictitious magnetic field produced by a guided light field may have a nonzero component $\mathbf{B}^{\mathrm{fict}}_{\perp}=
\mathbf{B}^{\mathrm{fict}}_x+\mathbf{B}^{\mathrm{fict}}_y=\mathbf{B}^{\mathrm{fict}}_r+\mathbf{B}^{\mathrm{fict}}_\varphi$ in the transverse plane as opposed to the fictitious magnetic field created by a paraxial light field.
\begin{figure}
\begin{center}
  \includegraphics{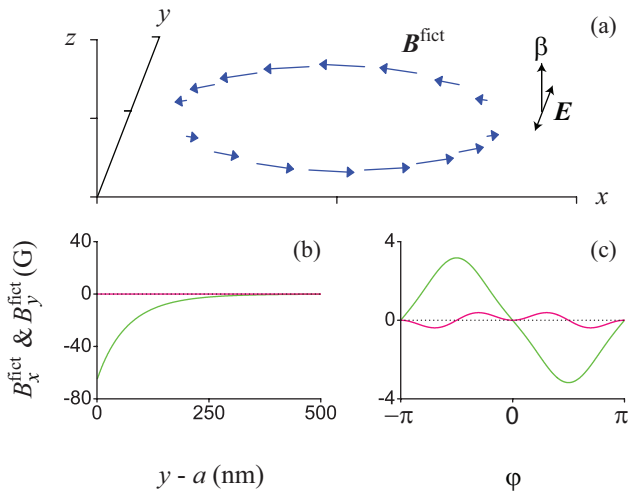}
 \end{center}
\caption{(a) Azimuthal vector profile (blue arrows) and (b) radial and (c) azimuthal dependences
of the components $B^{\mathrm{fict}}_x$  (green curves) and $B^{\mathrm{fict}}_y$ (pink curves) 
of the fictitious magnetic field $\mathbf{B}^{\mathrm{fict}}$ produced by a quasi-linearly polarized running-wave guided light field. The fiber radius is $a=250$ nm. The  wavelength and power of light are $\lambda=780$ nm  and $P=25$ mW, respectively. 
The field propagates along the $+z$ direction and is polarized along the $y$ direction.
In (a) and (c), the distance to the fiber surface is $r-a=224$ nm. In (b), the azimuthal angle is $\varphi=\pi/2$ (corresponding to the positive part of the $y$ axis). Other coordinates are arbitrary if not specified.
}
\label{fig8}
\end{figure} 

For quasi-linearly polarized running-wave modes (\ref{n28}), we have $\mathrm{Im}(\mathcal{E}_r\mathcal{E}_\varphi^*)=0$, which leads to
\begin{equation}\label{n33}
\mathbf{B}^{\mathrm{fict}}_{\mathrm{lin}}
=\frac{\alpha^v_{nJ}}{4\mu_Bg_{nJ}J}[\mathrm{Im}(\mathcal{E}_\varphi\mathcal{E}_z^*)\hat{\mathbf{r}}
+\mathrm{Im}(\mathcal{E}_z\mathcal{E}_r^*)\hat{\boldsymbol{\varphi}}].
\end{equation}
In this case, $\mathbf{B}^{\mathrm{fict}}$ lies in the fiber transverse plane $(x,y)$ (see Fig. \ref{fig8}).
It is clear from Eqs. (\ref{n28}) and (\ref{n33}) that the orientation direction of 
$\mathbf{B}^{\mathrm{fict}}$ reverses when we change the propagation direction $f$ of the light field.
We note that, for the azimuthal coordinate $\varphi=\varphi_0\pm\pi/2$, we have $\mathcal{E}_z=0$, which leads to the exact linear polarization and, consequently, to $\mathbf{B}^{\mathrm{fict}}=0$, that is, to the vanishing of the vector Stark shift~\cite{Lacroute12}.
Without loss of generality, we choose $\varphi_0=\pi/2$.
For this particular choice, we can show that
\begin{eqnarray}\label{n39}
\lefteqn{\mathbf{B}^{\mathrm{fict}}_{\mathrm{lin}}  
\propto |A|^2 \alpha^v_{nJ} K_1(qr)\sin\varphi\{\hat{\mathbf{x}}[(1-s)K_0(qr)}
\nonumber\\&&\mbox{}
-(1+s)K_2(qr)\cos2\varphi]
-\hat{\mathbf{y}}(1+s)K_2(qr)\sin2\varphi\}. \qquad
\end{eqnarray}
Since $s\simeq  -1$ and $K_0(qr)>K_2(qr)>0$, we have $|B_x^{\mathrm{fict}}|\gg |B_y^{\mathrm{fict}}|$. Thus,
the fictitious magnetic field $\mathbf{B}^{\mathrm{fict}}$ is oriented mainly along the axis $x$, which is perpendicular to the axis $z$ of the fiber and to the principal axis $y$ of the polarization of the quasi-linearly polarized guided light field. The calculations in Fig.~\ref{fig8} show, that the magnitude of the fictitious magnetic field for $r-a\approx200$~nm (typical atom-surface separation for atoms in a two-color trap) reaches up to a few Gauss. This is comparable to the magnitude of real magnetic fields present in conventional magnetic traps for cold atoms. In contrast, the gradient of the fictitious magnetic field for the case in Fig.~\ref{fig8} goes beyond what is possible with most conventional magnetic traps, reaching about 40 000 G/cm. These large gradients of the fictitious magnetic field in conjunction with the steep potentials of optical near field traps can pronounce effects that are beyond what is expected from the adiabatic approximation.

For standing-wave modes with quasi-linear polarization (\ref{n30}), 
the polarization is linear at every point in space. In this case, the fictitious magnetic field vanishes, that is,
\begin{equation}\label{n35}
\mathbf{B}^{\mathrm{fict}}_{\mathrm{lin}\mbox{-}\mathrm{st}}=0.
\end{equation}

For completeness, we discuss other nanofiber-guided light field configurations and the associated fictitious magnetic fields in the appendix.

\section{Tweaking the state-depended potentials of a nanofiber-based two-color trap}
\label{sec:VetschMagnetic}

Inspired by the physics of magnetic traps for cold atoms, we study the effect of an external magnetic offset field on the two-color trap. As main results we find, that the degeneracy point in Fig.~\ref{fig6} can be removed and, thus, spin flips can be suppressed. Moreover, we show that the Zeeman-state-dependent displacement of the trapping potential depends on the magnitude and direction of the offset field. The option of microwave cooling in the displaced potentials is suggested and Franck-Condon factors are calculated.

\subsection{Two-color trap in the presence of a homogeneous magnetic offset field}
\label{sec:magnetic}

\begin{figure}
\begin{center}
 \includegraphics{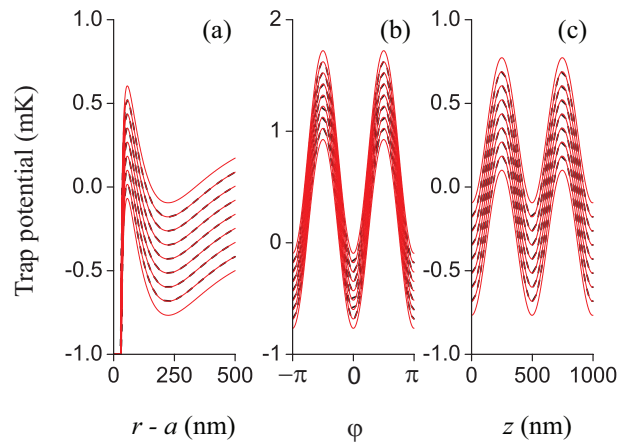}
 \end{center}
\caption{Radial (a), azimuthal (b), and axial (c) dependences of the potentials of 
the ground-state hfs levels $6S_{1/2}\,F=3$ (dashed black curves) and $6S_{1/2}\,F=4$ (solid red curves)
of a cesium atom in the nanofiber-based trap with a homogeneous magnetic field oriented along the $z$ axis.
The magnitude of the offset magnetic field is $B=5$ G.
The coordinates of the atom are $\varphi=0$ and $z=0$ in (a), $r-a=224$ nm and $z=0$ in (b),
and $r-a=224$ nm and $\varphi=0$ in (c). For parameters of the trap, see the text.
}
\label{fig9}
\end{figure}

\begin{figure}
\begin{center}
 \includegraphics{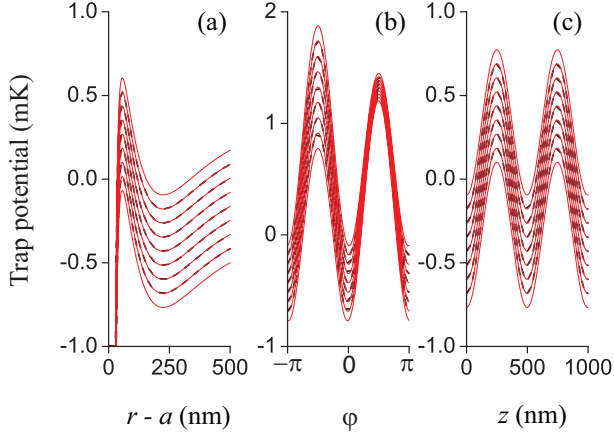}
 \end{center}
\caption{Same as Fig. \ref{fig9} but the applied magnetic field is oriented along the $x$ axis.
}
\label{fig10}
\end{figure}

Let us consider the effect of a weak external magnetic field $\mathbf{B}$ on the ground-state potential of the atom in the two-color nanofiber-based trap (see Fig. \ref{fig2}). We calculate numerically the adiabatic potential of the atom in the presence of the magnetic field using the formalism outlined in Sec.~\ref{sec:theory}. We plot in Figs. \ref{fig9} and \ref{fig10} the potentials of the ground-state hfs levels $6S_{1/2}\,F=3$ and $6S_{1/2}\,F=4$ of a cesium atom in the nanofiber-based trap with a homogeneous magnetic field oriented along the $z$ and $x$ axes, respectively. The magnitude of the magnetic field is $B=5$ G.

The figures show that the degeneracy of  the Zeeman sublevels is lifted in all directions, as opposed to the situation of the red curves in Figs. \ref{fig5}--\ref{fig7}, by the light-induced fictitious magnetic field $\mathbf{B}^{\mathrm{fict}}$ 
and the applied magnetic field $\mathbf{B}$, that is, by the total effective (fictitious $+$ real) magnetic field $\mathbf{B}^{\mathrm{total}}=\mathbf{B}^{\mathrm{fict}}+\mathbf{B}$. 
The splittings of the Zeeman sublevels of the ground state depend on the magnitudes of the components $\mathbf{B}^{\mathrm{fict}}$ and $\mathbf{B}$ and on their relative orientation. It is clear that the differential shifts of the sublevels vary in space. 
Note that, unlike Fig. \ref{fig9}(b), the two peaks in Fig. \ref{fig10}(b) are not symmetric.
The difference between the peaks in Fig. \ref{fig10}(b) is due to the fact that, according to Figs. \ref{fig8}(a) and \ref{fig8}(c), at the peak positions $\varphi=-\pi/2$ and $\varphi=\pi/2$, the vector $\mathbf{B}^{\mathrm{fict}}$ is oriented along the direction $\hat{\mathbf{x}}$ (parallel to
$\mathbf{B}$) and $-\hat{\mathbf{x}}$ (opposite to $\mathbf{B}$), respectively, leading to differing magnitudes of the total effective magnetic field $\mathbf{B}^{\mathrm{total}}$ and, consequently, to differing Zeeman-state shifts.
Note that Figs. \ref{fig10}(a) and \ref{fig10}(c) are identical to Figs. \ref{fig9}(a) and \ref{fig9}(c), respectively.
The reason is that, for $\varphi=0,\pm\pi$, that is, along the $x$ axis, the fictitious magnetic field $\mathbf{B}^{\mathrm{fict}}$ vanishes and, therefore, 
the energy shifts of the sublevels of the ground state is determined only by the scalar polarizability and the applied magnetic field $\mathbf{B}$. Such shifts do not depend on the direction of $\mathbf{B}$.

\begin{figure}
\begin{center}
 \includegraphics{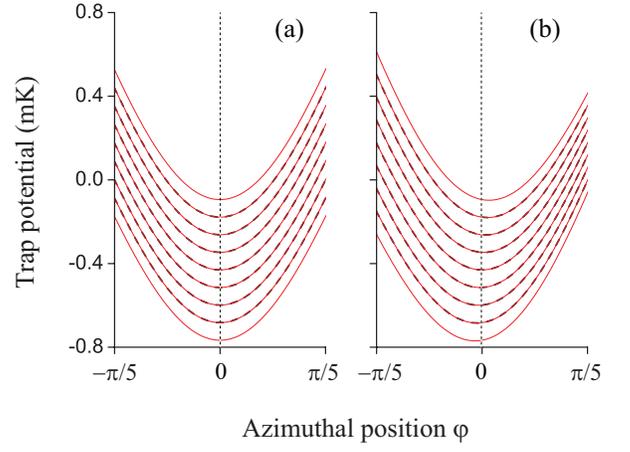}
 \end{center}
\caption{Expanded view of Figs. \ref{fig9}(b) and \ref{fig10}(b).
}
\label{fig11}
\end{figure}

Figure \ref{fig11} shows clearly that 
the residual degeneracy of the ground-state sublevels, observed  at the points with $\varphi=0$ or $\pm\pi$ in the situation of Fig. \ref{fig6}(b), is lifted
in the situations of Figs. \ref{fig9}(b) and \ref{fig10}(b). This is due to the applied magnetic field. When the splitting between the Zeeman sublevels is large enough, the atom can adiabatically follow a given trapping potential thereby maintaining its $M$ state. 

We observe from Fig. \ref{fig11}(b) that, when the external magnetic field is oriented along the $x$ direction, 
the minima of the state-dependent trapping potentials are
displaced from each other in the azimuthal direction.
In order to get deep insight into the displacement of the potentials, we consider the close vicinity of the central trap position $(r=r_m,\varphi=0)$. In this region, the component $B^{\mathrm{fict}}_x$ of the fictitious magnetic field $\mathbf{B}^{\mathrm{fict}}$ can be considered as a linear function of $\varphi$ while the component $B^{\mathrm{fict}}_y$ can be neglected [see Fig. \ref{fig8}(c) and Eq. (\ref{n39})]. Since the applied magnetic field $\mathbf{B}$ in the case of Figs. \ref{fig10} and \ref{fig11} is oriented along the $x$ direction and is significant enough, the fictitious magnetic field
$\mathbf{B}^{\mathrm{tot}}$ is oriented mainly along the $x$ direction and can therefore be approximated by the $x$ component
\begin{equation}\label{n46a}
B^{\mathrm{tot}}_x=B+B^{\mathrm{fict}}_x \simeq  B+B'_{\mathrm{fict}} \varphi. 
\end{equation}
Here the parameter $B'_{\mathrm{fict}}$ is defined as $B'_{\mathrm{fict}}=(\partial B^{\mathrm{fict}}_x /\partial\varphi)|_{\varphi=0}$. 
Hence, the azimuthal potential of a sublevel $|FM\rangle$ of the ground state can be written as  $U_{FM}(\varphi)\equiv U(\varphi)=U_s(\varphi)+\mu_Bg_{nJF}M(B+B'_{\mathrm{fict}} \varphi)$, where $U_s(\varphi)$ is the scalar Stark shift. In the close vicinity of the trap minimum, the scalar shift can be approximated as $U_s(\varphi)\simeq (1/2)U_s''\varphi^2+\mathrm{const}$, where $U_s''=(\partial^2 U_s(\varphi) /\partial\varphi^2)|_{\varphi=0}$. Then, we obtain the azimuthal potential $U_{FM}(\varphi)=(1/2)U_s''\varphi^2+\mu_Bg_{nJF} MB'_{\mathrm{fict}}\varphi+\mathrm{const}$, which is the potential of a displaced harmonic oscillator. The equilibrium position of this potential is displaced from the central trap position $(r=r_m,\varphi=0)$ in the azimuthal direction by the angle $\Delta\varphi_{FM}=-\mu_Bg_{nJF} MB'_{\mathrm{fict}}/U_s''$. Thus, the displacement $\Delta\varphi_{FM}$ is proportional to the quantum number $M$, amounting to a relative displacement of $0.01$~rad or about $4.7$~nm for $\Delta M=1$.

\subsection{Franck-Condon factors}
When we apply a microwave field near to resonance with the hyperfine splitting of the ground state ($\sim9.2$ GHz for atomic cesium), the atom can change from a state $|F'M'\nu'\rangle\equiv|F'M'\rangle|\nu'\rangle$ of the atom in the hfs sublevel $|F'=4,M'\rangle$ to a state $|FM\nu\rangle\equiv|FM\rangle|\nu\rangle$ of the atom in the hfs sublevel $|F=3,M\rangle$ and vice versa through a magnetic dipole transition. Here, $\nu'$ and $\nu$  are quantum numbers for the center-of-mass motion of the atom in the trapping potentials of the atomic internal states $|F'M'\rangle$ and $|FM\rangle$, respectively, and  $|\nu'\rangle$ and $|\nu\rangle$ are the corresponding center-of-mass eigenfunctions. Since the magnetic dipole operator acts only on the internal states, the transition probability depends on the
Franck-Condon factor $F_{\nu\nu'}\equiv F_{(FM)\nu(F'M')\nu'} \equiv|\langle\nu|\nu'\rangle|^2$, which characterizes the spatial overlap between the atomic center-of-mass wave functions $|\nu\rangle$ and $|\nu'\rangle$.

\begin{figure}
\begin{center}
 \includegraphics{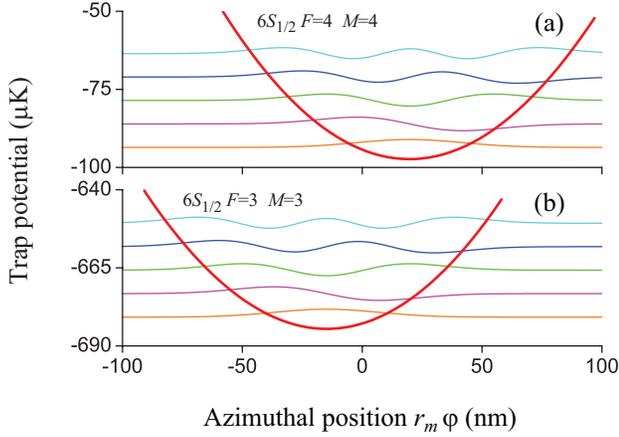}
 \end{center}
\caption{Eigenfunctions (thin lines) for the center-of-mass motion of a cesium atom in the azimuthal potentials (thick red lines) of the sublevels $6S_{1/2}F=4,M=4$ (a) and $6S_{1/2}F=3,M=3$ (b) in the vicinity of the trap minimum at $r_m-a=224$ nm. 
The offset magnetic field is $B=5$ G and is aligned along the $x$ axis. For parameters of the trap, see the text.
}
\label{fig12}
\end{figure}

\begin{figure}
\begin{center}
 \includegraphics{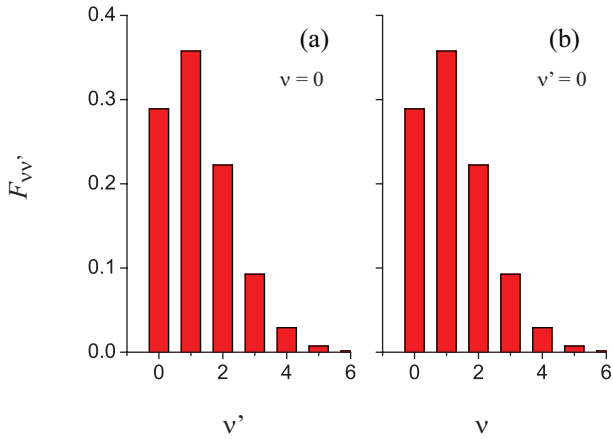}
 \end{center}
\caption{
Franck-Condon factor $F_{\nu\nu'}$ for 
the overlap between the motional states $|\nu\rangle$ and $|\nu'\rangle$ of the Zeeman sublevels $|F=3,M=3\rangle$ and $|F'=4,M'=4\rangle$, respectively, of the cesium ground state $6S_{1/2}$. The offset magnetic field is $B=5$ G and is aligned along the $x$ axis. For parameters of the trap, see the text.
}
\label{fig13}
\end{figure}

In the absence of the vector Stark shift, the trapping potentials of the sublevels $|F'M'\rangle$ and $|FM\rangle$ of the hfs levels of the ground state are identical to each other because they are determined by the scalar Stark shift, which is the same for all of these sublevels. In this case, we have $\langle\nu|\nu'\rangle=\delta_{\nu\nu'}$, which leads to $F_{\nu\nu'}=\delta_{\nu\nu'}$. This formula indicates that, in the absence of the vector Stark shift, a microwave transition between the sublevels $|F'M'\rangle$ and $|FM\rangle$ of the ground state cannot lead to a change in the motional state. In the presence of the vector Stark shift, the trapping potentials of the sublevels $|F'M'\rangle$ and $|FM\rangle$ of the hfs levels $F'=I+1/2$ and $F=I-1/2$, respectively, of the ground state are not identical to each other anymore if $M'+M\not=0$. They are split from each other by an integer multiple $M'+M$ of the quantity $\mu_Bg_{nJ}B^{\mathrm{total}}/(2I+1)$, which 
varies in space. 
When the applied magnetic field $\mathbf{B}$ is oriented in the $x$ direction, 
the state-dependent trapping potentials are, as shown in the previous subsection, displaced from each other in the azimuthal direction. 
In this case, a microwave transition between the atomic electronic states $|F'M'\rangle$ and $|FM\rangle$ of two different hfs levels $F'$ and $F$ the ground state may lead to a change in the atomic center-of-mass motional state.

It is computationally difficult to calculate the Franck-Condon factors between the states of the center-of-mass motion of the atom in a three-dimensional trapping potential. Therefore, we limit ourselves to the one-dimensional motion of the atom along the azimuthal direction. To be specific, we consider the motion of the atom along the azimuthal line $\varphi$ going through the trap minimum at $r_m-a\simeq  224$ nm in the situation of Figs. \ref{fig10}(b) and \ref{fig11}(b). We plot in Fig. \ref{fig12} the wave functions of the eigenstates of a cesium atom in the potentials of the electronic sublevels $6S_{1/2}F=4,M=4$ and $6S_{1/2}F=3,M=3$. Our numerical calculations show that the trap frequencies in the azimuthal direction are about 157 kHz for both the upper and lower hfs levels of the ground state. The characteristic size of the ground motional states of the potentials is about 16 nm which is comparable to the displacements of the azimuthal potential minima for different Zeeman states.

We plot in Fig. \ref{fig13} the Franck-Condon factor $F_{\nu\nu'}$ as functions of $\nu'$ and $\nu$.
We observe that $F_{\nu\nu'}\not=0$ for a number of transitions with $\nu\not=\nu'$. This indicates the possibility of a change in the motional state when the atom changes its hfs state in a microwave transition. Comparison between parts (a) and (b) of Fig. \ref{fig13} shows that $F_{\nu\nu'}\simeq  F_{\nu'\nu}$, that is,
$F_{(FM)\nu(F'M')\nu'}\simeq  F_{(FM)\nu'(F'M')\nu}$. The reason is that, in the close vicinity of the trapping minimum, the potentials $U_{FM}(\varphi)$ for different sublevels $|FM\rangle$ have approximately the same form but are displaced in the $\varphi$ direction. 

We note that along the $x$ axis, the fictitious magnetic field is zero. The one-dimensional potentials $U_{FM}$ of different sublevels $|FM\rangle$ along these directions are identical up to a constant, namely $U_{FM}=U_s+\mu_Bg_{nJF}MB$. This leads to $F_{\nu\nu'}=\delta_{\nu\nu'}$.

\section{Summary}
\label{sec:summary}
We have calculated the adiabatic trapping potentials of cesium atoms in a nanofiber-based two-color trap with and without an externally applied static magnetic field. This external field can simply be added to the fictitious magnetic field which is produced by the combined action of the vector polarizability and the ellipticity of the polarization of the guided light field. In the case of optical nanofibers, the ellipticity and intensity of the guided field and, consequently, the fictitious magnetic field varies on a nanoscale. We showed that this results in a remarkably high magnetic field gradient on the order of several G/$\mu$m for the nanofiber-based two-color trap in~\cite{Rauschenbeutel10}. This leads to a Stern-Gerlach-type mechanical force on the trapped atoms which, in the absence of an external magnetic field, mutually displaces the adiabatic potentials of different Zeeman sublevels $|FM\rangle$ of the atomic ground state. 

The quantitative understanding of the light-induced fictitious magnetic field allowed us to devise and to analyze trap configurations that make use of a magnetic offset field in order to suppress the state-dependent displacement of the potentials and / or to overcome its negative effects on the coherence of nanofiber-trapped atoms~\cite{Lacroute12,Reitz13}. Moreover, our work suggests that it is possible to employ the fictitious magnetic field for tailoring trapping potentials and for controlling and manipulating the internal and external atomic states. As an example, we show that for an appropriately applied external magnetic field, the state dependent potentials result in nonzero Franck-Condon factors for the overlap between different motional states of different hfs levels of the electronic ground state of the trapped atoms. This then opens the possibility of microwave cooling of nanofiber-trapped atoms.

\begin{acknowledgments}
We thank R.~Grimm and H.~J.~Kimble for helpful discussions. 
Financial support by the Wolfgang Pauli Institute and the Austrian Science Fund (FWF; Lise Meitner project No.~M 1501-N27 and SFB NextLite project No.~F 4908-N23) is gratefully acknowledged.
\end{acknowledgments}

\appendix
\section{Fictitious magnetic fields of nanofiber-guided light fields -- continued}
In general, an arbitrary guided mode of a single-mode fiber at a given frequency can be composed as a superposition of HE$_{11}$ modes with different polarizations and propagation directions. To complete the discussion of fictitious magnetic fields generated by fiber-guided light fields as started in the main text, we present results for the following light field configurations: (a) a quasi-circularly polarized running-wave mode, 
(b) a pair of counter-propagating modes with the same quasi-circular polarization, 
(c) a pair of counter-propagating modes with the opposite quasi-circular polarization, 
and (d) a pair of counter-propagating modes with opposite quasi-linear polarization. A combination of (a) and (b) would form a nanofiber-based trap with toroidal trapping potential whereas a combination of (a) and (c) is relevant when realizing a nanofiber-based helical trap~\cite{Reitz12}.

(a) The electric component of the guided light field in a quasi-circularly polarized running-wave mode is given by
\begin{equation}
\boldsymbol{\mathcal{E}}_{\mathrm{circ}} 
= A(\hat{\mathbf{r}}e_r+l\hat{\boldsymbol{\varphi}}e_\varphi+
f\hat{\mathbf{z}}e_z) e^{if\beta z +il\varphi}.
\label{n26}
\end{equation}
The index $l=1$ or $-1$ refers to the left- or
right-handed circulation, respectively, of the transverse component of the field around the fiber axis.
The index $f=1$ or $-1$ stands for the forward or backward propagation direction, respectively.
The coefficient $A$ is determined from the propagating power of the light field in the axial direction.
\begin{figure}
\begin{center}
  \includegraphics{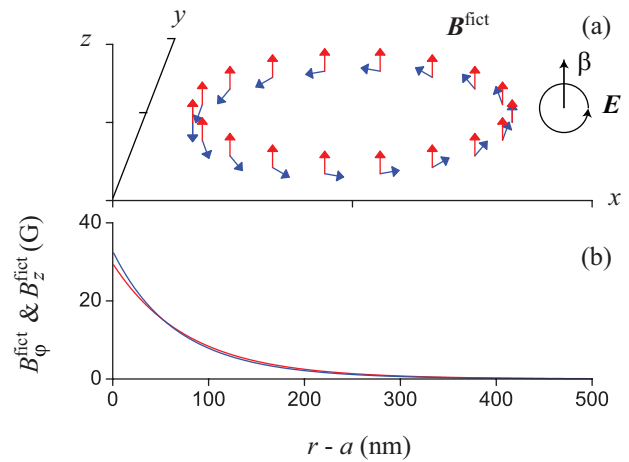}
 \end{center}
\caption{Azimuthal vector profile (a) and radial dependence (b) of the tangential component 
$B^{\mathrm{fict}}_\varphi$ (blue arrows and blue curve) and 
the axial component $B^{\mathrm{fict}}_z$ (red arrows and red curve)
of the fictitious magnetic field produced by a quasi-circularly polarized running-wave guided light field. 
The field propagates along the $+z$ direction and is counterclockwise polarized.
The fiber radius is $a=250$ nm. 
The  wavelength and power of light are $\lambda=780$ nm  and $P=25$ mW, respectively.
In (a), the distance to the fiber surface is $r-a=224$ nm. Other coordinates are arbitrary if not specified.
}
\label{fig14}
\end{figure} 

According to Eqs. (\ref{n25}), the function $e_r$ is imaginary-valued and the functions $e_\varphi$ and $e_z$ are real-valued.
Therefore, for quasi-circularly polarized running-wave modes (\ref{n26}), we have $\mathrm{Im}(\mathcal{E}_\varphi\mathcal{E}_z^*)=0$, and consequently
the fictitious magnetic field $\mathbf{B}^{\mathrm{fict}}$, given by Eq. (\ref{n20}), reduces to
\begin{equation}\label{n32}
\mathbf{B}^{\mathrm{fict}}_{\mathrm{circ}}
=\frac{\alpha^v_{nJ}}{4\mu_Bg_{nJ}J}[ 
\mathrm{Im}(\mathcal{E}_z\mathcal{E}_r^*)\hat{\boldsymbol{\varphi}}+\mathrm{Im}(\mathcal{E}_r\mathcal{E}_\varphi^*)\hat{\mathbf{z}}].
\end{equation}
In this case, $\mathbf{B}^{\mathrm{fict}}$ has a tangential component $B^{\mathrm{fict}}_\varphi$ 
and an axial component $B^{\mathrm{fict}}_z$ (see Fig. \ref{fig14}).
It is clear from Eqs. (\ref{n26}) and (\ref{n32}) that the orientation direction $\pm \hat{\boldsymbol{\varphi}}$ of the tangential component $B^{\mathrm{fict}}_\varphi$ depends on the propagation direction $f=\pm 1$
but not on the polarization circulation direction $l=\pm1$, while
the orientation direction $\pm \hat{\mathbf{z}}$ of the axial component 
$B^{\mathrm{fict}}_z$ depends on the polarization circulation direction $l=\pm1$ but not on
the propagation direction $f=\pm 1$. 

\begin{figure}
\begin{center}
  \includegraphics{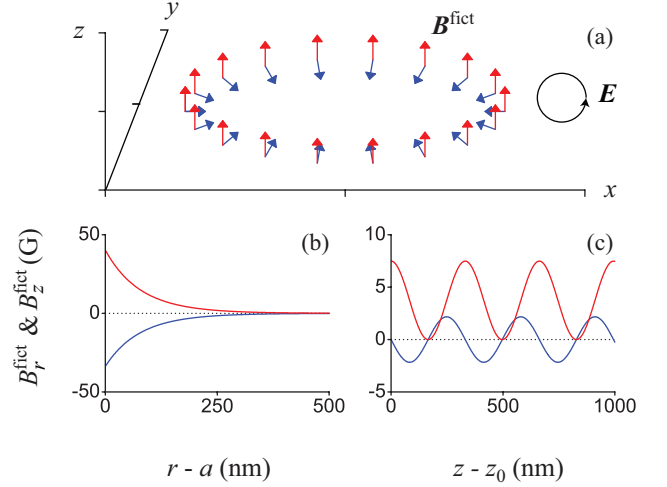}
 \end{center}
\caption{Azimuthal vector profile (a) and radial (b) and axial (c) dependences
of the components $B^{\mathrm{fict}}_r$  (blue arrows and curves) and $B^{\mathrm{fict}}_z$ (red arrows and curves) 
of the fictitious magnetic field $\mathbf{B}^{\mathrm{fict}}$ produced by a quasi-circularly polarized standing-wave guided light field. 
The field is counterclockwise polarized.
The fiber radius is $a=250$ nm. The  wavelength and power of light are $\lambda=780$ nm  and $P=2\times 25$ mW, respectively.
The radial position is $r-a=224$ nm in (a) and (c).
The axial position is $z-z_0=100$ nm in (a) and (b). Other coordinates are arbitrary if not specified.
}
\label{fig15}
\end{figure}

(b) For a pair of counter-propagating modes with the same quasi-circular polarization, the electric field is given by
\begin{eqnarray}
\boldsymbol{\mathcal{E}}_{\mathrm{circ}\mbox{-}\mathrm{st}}
&=&
A[(\hat{\mathbf{r}}e_r  
+l\hat{\boldsymbol{\varphi}}e_\varphi)\cos\beta (z-z_0)
\nonumber\\&&\mbox{}
+i\hat{\mathbf{z}}e_z\sin\beta (z-z_0)]e^{il\varphi}.
\label{n29}
\end{eqnarray}
In this case, we have $\mathrm{Im}(\mathcal{E}_z\mathcal{E}_r^*)=0$, which leads to
\begin{equation}\label{n34}
\mathbf{B}^{\mathrm{fict}}_{\mathrm{circ}\mbox{-}\mathrm{st}}
=\frac{\alpha^v_{nJ}}{4\mu_Bg_{nJ}J}[\mathrm{Im}(\mathcal{E}_\varphi\mathcal{E}_z^*)\hat{\mathbf{r}}
+\mathrm{Im}(\mathcal{E}_r\mathcal{E}_\varphi^*)\hat{\mathbf{z}}].
\end{equation}
The resulting $\mathbf{B}^{\mathrm{fict}}$ thus has a radial component $B^{\mathrm{fict}}_r$ 
and an axial component $B^{\mathrm{fict}}_z$ (see Fig. \ref{fig15}).
Equations (\ref{n29}) and (\ref{n34}) and Fig. \ref{fig15}(c) show that the magnitudes of 
$B^{\mathrm{fict}}_r$ and $B^{\mathrm{fict}}_z$ oscillate when we vary the axial position $z$.
The modulations of $B^{\mathrm{fict}}_r$ and  $B^{\mathrm{fict}}_z$ along the fiber axis are governed by the functions $\cos\beta (z-z_0)\sin\beta (z-z_0)$ 
and $\cos^2\beta (z-z_0)$, respectively. 
Hence, the sign of the radial component $B^{\mathrm{fict}}_r$ varies with varying the axial position $z$ while 
the sign of the axial component $B^{\mathrm{fict}}_z$ does not.
Moreover, at the axial quasi-node positions $\beta (z-z_0)=(n+1/2)\pi$, where $n$ is an arbitrary integer,
both components $B^{\mathrm{fict}}_r$ and $B^{\mathrm{fict}}_z$ and consequently the fictitious magnetic field  $\mathbf{B}^{\mathrm{fict}}$  
become zero everywhere in the fiber transverse plane. At these specific axial positions, the transverse component $\mathcal{E}_\perp$
of the electric field $\boldsymbol{\mathcal{E}}$
vanishes, that is, the field $\boldsymbol{\mathcal{E}}$ becomes linearly polarized along the fiber axis $z$.
We note that the signs of the components $B^{\mathrm{fict}}_r$ and  $B^{\mathrm{fict}}_z$ depend on the polarization circulation direction $l=\pm1$.

(c) For a pair of counter-propagating modes with opposite quasi-circular polarizations, the field is
\begin{eqnarray}
\boldsymbol{\mathcal{E}}_{\sigma^l\mbox{-}\sigma^{-l}}
&=&A\{\hat{\mathbf{r}}e_r \cos[\beta z +l(\varphi-\varphi_0)]
\nonumber\\&&\mbox{}
+il\hat{\boldsymbol{\varphi}}e_\varphi \sin[\beta z +l(\varphi-\varphi_0)]
\nonumber\\&&\mbox{}
+i\hat{\mathbf{z}}e_z \sin[\beta z +l(\varphi-\varphi_0)]\}.
\label{n29a}
\end{eqnarray}
In this case, the polarization is linear at every point in space. In this case, the fictitious magnetic field vanishes, that is,
\begin{equation}\label{n35a}
\mathbf{B}^{\mathrm{fict}}_{\sigma^l\mbox{-}\sigma^{-l}}=0.
\end{equation}

\begin{figure}
\begin{center}
  \includegraphics{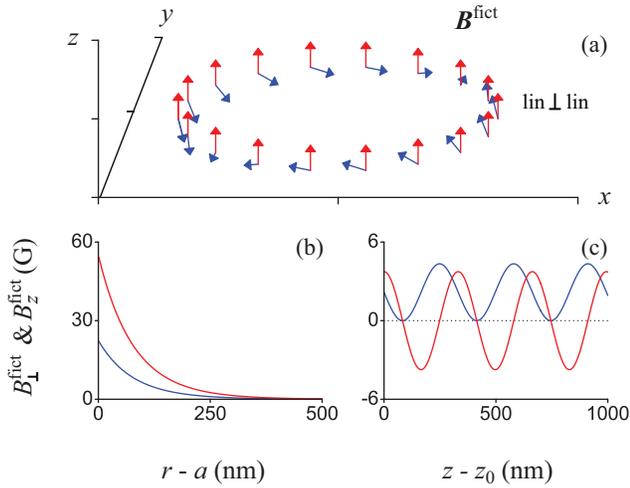}
 \end{center}
\caption{Azimuthal vector profile (a) and radial (b) and axial (c) dependences
of the transverse component $B^{\mathrm{fict}}_{\perp}$ (blue arrows and curves) and the axial component $B^{\mathrm{fict}}_z$ (red arrows and curves) of the fictitious magnetic field $\mathbf{B}^{\mathrm{fict}}$ produced by a pair of counter-propagating guided light fields with orthogonal quasi-linear polarizations. The parameters $z_0=0$ and $\varphi_0=0$ are chosen.
The fiber radius is $a=250$ nm. The  wavelength and power of light are $\lambda=780$ nm  and $P=2\times 25$ mW, respectively.
The radial position is $r-a=224$ nm in (a) and (c). The azimuthal position is $\varphi=\pi/4$ nm in (b) and (c).
The axial position is $z=20$ nm in (a) and $z=0$ in (b). Other coordinates are arbitrary if not specified.
}
\label{fig16}
\end{figure}

(d) For a pair of counter-propagating modes with opposite quasi-linear polarizations, the field is
\begin{eqnarray}
\boldsymbol{\mathcal{E}}_{\mathrm{lin}\perp\mathrm{lin}}
&=&A\Big\{\hat{\mathbf{r}}e_r \{i\sin[\beta (z-z_0)+\varphi-\varphi_0]
\nonumber\\&&\mbox{}\qquad
+\cos[\beta(z-z_0)-\varphi+\varphi_0]\}
\nonumber\\&&\mbox{}
+\hat{\boldsymbol{\varphi}}e_\varphi\{\cos[\beta (z-z_0)+\varphi-\varphi_0]
\nonumber\\&&\mbox{}\qquad
-i\sin[\beta(z-z_0)-\varphi+\varphi_0]\}
\nonumber\\&&\mbox{}
+\hat{\mathbf{z}}e_z \{\cos[\beta (z-z_0)+\varphi-\varphi_0]
\nonumber\\&&\mbox{}\qquad
+i\sin [\beta(z-z_0)-\varphi+\varphi_0]\}\Big\}.
\nonumber\\
\label{n30a}
\end{eqnarray}
In this case, the radial, azimuthal, and axial components of the fictitious magnetic field are, in general, nonzero and are given by the expression
\begin{eqnarray}\label{n35b}
\mathbf{B}^{\mathrm{fict}}_{\mathrm{lin}\perp\mathrm{lin}}
&=&\frac{\alpha^v_{nJ}}{4\mu_Bg_{nJ}J}[ 
\mathrm{Im}(\mathcal{E}_\varphi\mathcal{E}_z^*)\hat{\mathbf{r}}
+\mathrm{Im}(\mathcal{E}_z\mathcal{E}_r^*)\hat{\boldsymbol{\varphi}}
\nonumber\\&&\mbox{}
+\mathrm{Im}(\mathcal{E}_r\mathcal{E}_\varphi^*)\hat{\mathbf{z}}].
\end{eqnarray}
It is clear from the above expression and from Fig. \ref{fig16} that both the axial component $B^{\mathrm{fict}}_z$ and the transverse component $B^{\mathrm{fict}}_{\perp}=\sqrt{|B^{\mathrm{fict}}_x|^2+|B^{\mathrm{fict}}_y|^2}$ of the fictitious magnetic field can be nonzero. Figure \ref{fig16} shows that the magnitude and orientation of the fictitious magnetic field vary in space in a complicated manner. We observe from Fig. \ref{fig16} that both the axial component $B^{\mathrm{fict}}_z$ and the transverse component $B^{\mathrm{fict}}_{\perp}$ become zero along some specific radial directions at some specific axial positions.
A close inspection of Eqs. (\ref{n30a}) and (\ref{n35b}) reveals that $B^{\mathrm{fict}}$ vanishes at the points with the coordinates 
$\varphi-\varphi_0=\pi/4+n\pi/2$ and $\beta (z-z_0)=\pi/4+m\pi/2$, where $n$ and $m$ are arbitrary integer numbers.

The above results show that one can generate some specific configurations of the fictitious magnetic field $\mathbf{B}^{\mathrm{fict}}$ by using an appropriate guided light field. This opens up possibilities for controlling and manipulating the fictitious magnetic field on the nanoscale, particularly when using the so-called tune-out wavelengths~\cite{Arora11} that impose no scalar but only a vector light shift.

\end{document}